\documentclass[aps,showpacs,twocolumn,floatfix]{revtex4}
\usepackage{amsmath}
\usepackage{amssymb}
\usepackage[dvips]{graphics,graphicx}
\begin{document}
\title{Bloch oscillations of Bose-Einstein condensates: Quantum counterpart of dynamical instability}
\author{Andrey R. Kolovsky, }
\affiliation{Kirensky Institute of Physics and
Siberian Federal University, 660036 Krasnoyarsk, Russia}
\author{Hans J\"urgen Korsch}
\affiliation{Fachbereich Physik, Technische Universit{\"a}t Kaiserslautern,
D-67653 Kaiserslautern, Germany}
\author{Eva-Maria Graefe}
\affiliation{Fachbereich Physik, Technische Universit{\"a}t Kaiserslautern,
D-67653 Kaiserslautern, Germany}
\date{\today}

\begin{abstract}
We study the Bloch dynamics of a quasi one-dimensional Bose-Einstein 
condensate of cold atoms in a tilted optical lattice modeled by
a Hamiltonian of Bose-Hubbard type: The corresponding mean-field system 
described by a discrete nonlinear Schr\"odinger equation can exhibit 
dynamical (or modulation) instability due to chaotic dynamics and 
equipartition over the quasimomentum modes. It is shown that these 
phenomena are related to Bogoliubov's depletion  
of the  Bose-Einstein condensate and a decoherence of the  
condensate in the many-particle description.
Three types of dynamics are distinguished: (i) decaying oscillations 
in the region of dynamical instability, and (ii) persisting Bloch 
oscillations or (iii) periodic decay and revivals in the region of 
stability.
\end{abstract}
\maketitle


\section{Introduction}
Recently, considerable attention has been paid to the dynamics of 
cold atoms and Bose-Einstein condensates (BECs) loaded
into an optical lattice with a static, e.g., gravitational, force (see \cite{Daha96,Raiz97} and \cite{Ande98,Mors01,Cris02,Jona03,Ott04,04bloch_bec,Wimb05,Sias07,Gust08a,Fatt08}). 
In the limit of vanishing particle interaction the system shows 
Bloch oscillations (BO) (for a review see, e.g., \cite{04bloch}). 
However, it is known that the interaction between the atoms leads to 
modifications and possibly even to a breakdown of these Bloch oscillations.

The most prominent theoretical approach to investigate the dynamics 
is the reduction of the many-particle system to a mean-field
description via the (nonlinear) Gross-Pitaevskii equation. 
The description can be further simplified by 
discretizing it in terms of Wannier functions localized on the
lattice sites. In the many-particle description this yields a Bose-Hubbard model and 
accordingly the mean-field dynamics is described by the discrete nonlinear Schr\"odinger
equation (DNLSE) (see, e.g., \cite{Smer03}). 
It should be pointed out that the mean-field approximation 
is formally equivalent to a classical limit for single particle quantum mechanics.
Therefore, it is often denoted as `(pseudo)classical', although it still describes 
a quantum system. This becomes evident in the limit of vanishing interaction, 
where it reduces to a single particle Schr\"odinger equation. Nevertheless, the 
formal similarity of the many-particle to mean-field transition and the quantum 
classical correspondence allows for the 
application of semiclassical methods as well as the  
investigation of topics such as quantum chaos within the framework of 
ultracold atoms \cite{Holt01a,Moss06,Wu06,07semiMP,Luo08,08kicked,Weis08}. 

The mean-field system of interest in the present study shows 
a rich structure of mixed and chaotic behavior and our main aim will be 
to identify the counterparts of some pronounced features in the 
corresponding many-particle system. For this purpose we compare the DNLSE 
dynamics with the underlying microscopic many-particle system, described by 
the 1D bosonic Hubbard model with the Hamiltonian
\begin{equation}
\label{BHham}
\widehat{H}= \sum_l \epsilon_l \hat n_l 
-\frac{J}{2}\sum_l\Big(\hat{a}^\dag_{l+1}\hat{a}_l+h.c.\Big)
+\frac{W}{2}\sum_l \hat{n}_l(\hat{n}_l-1) \;,
\end{equation}
where $\hat{a}^\dag_l$ and $\hat{a}_l$ are bosonic creation and annihilation 
operators for the $l$th lattice site, and $\hat{n}_l=\hat{a}^\dag_l\hat{a}_l$ are 
the associated number operators.
The hopping energy and the particle interaction are denoted by $J$ and
$W$, respectively, and $\epsilon_l=dFl$ is the on-site energy where $d$
is the lattice period and $F$ the magnitude of a static force.
Experimentally, all these quantities can be controlled separately. However, the 
validity of the Bose-Hubbard model is based on certain assumptions such as the 
single band tight binding approximation, which may break down in some parameter 
regions. Yet, for the following studies only the ratios of the quantities are of 
interest for the qualitative behavior of the system which offers an additional 
freedom to maintain the validity of the Bose-Hubbard Hamiltonian. Therefore, we 
believe that the 
observations reported in the following should be experimentally accessible.

The Hamiltonian commutes with $\hat N=\sum_l\hat n_l$
and therefore the total number $N$ of particles is conserved. 
Here we consider the macroscopic limit $N\rightarrow \infty$, $W\rightarrow 0$ 
with constant $g=WN/L$ where the number of sites, $L$, is kept finite. 
In this limit the mean-field approximation can be applied, which usually is formulated 
as replacing the bosonic operators $\hat a_l$, $\hat a^\dagger_l$ by complex numbers 
$a_l$, $a^*_l$, the components of an effective single particle wave function which 
appear as `classical' canonical variables. The resulting mean-field Hamiltonian 
function is given by 
\begin{equation}
\label{meanfieldham}
H\!=\! \sum_l \epsilon_l |a_l|^2
\!-\!\frac{J}{2}\sum_l\Big(a^*_{l+1}\,a_l+c.c.\Big)
\!+\!\frac{g}{2}\sum_l |a_l|^4 \,,
\end{equation}
up to a term proportional to $\sum_l |a_l|^2$, which is an integral of motion. 
The DNLSE can be formulated via the canonical equations of motion
\begin{equation}
i\,\hbar \dot a_l=\partial H/\partial a^*_l\ ,\quad  
i\,\hbar \dot a_l^*=-\partial H/\partial a_l\,.
\label{caneq}
\end{equation}
One aim of the present paper is to investigate the validity of this mean-field 
approach for a system with finite particle number in the different parameter regions. 

For nonvanishing particle interaction $g\neq0$
the mean-field Bloch oscillations are damped \cite{04bloch_bec,Wimb05}. 
In this context the main phenomena are the dynamical instability which is also known 
as modulation instability (see, e.g., \cite{Kolo04a,Zhen04}) and the equipartition 
over the quasimomentum modes (also denoted
as thermalization) due to the onset of classical chaos in the DNLSE. 
On the other hand, within the many-particle approach the main
phenomenon induced by the interaction is a decay of Bloch oscillations due
to decoherence \cite{04bloch,04bloch_bec}. The present analysis shows a direct relation 
between these classical and quantum phenomena.
We argue that the quantum manifestations of both dynamical instability
and equipartition can be understood in terms of the depletion of the 
Floquet-Bogoliubov states, defined as the ``low-energy'' eigenstates of the evolution 
operator over one Bloch period. 

We furthermore go beyond the traditional single trajectory mean-field treatment and, 
following a recent suggestion \cite{07phase}, average the dynamics over an ensemble 
of trajectories given by the Husimi distribution of the initial many-particle state. 
It is shown that this method is capable of 
describing important features of the many-particle dynamics. 

The paper is organized as follows: In Sec.~\ref{sec2} we discuss the mean-field 
dynamics, in particular the stability properties of the Bloch oscillation and its 
relation to chaotic dynamics. The corresponding many-particle system is analyzed 
in Sec.~\ref{sec3}, mainly based on the Floquet-Bogoliubov states whose depletion 
properties provide a measure for the many-particle stability which can be compared 
to the mean-field behavior. We summarize our results and end with a short outlook 
in Sec.~\ref{sec-concl}.

\section{Mean-field dynamics}
\label{sec2}

Evaluating the canonical equations of motion (\ref{caneq}) with
the Hamiltonian function (\ref{meanfieldham}) yields the
mean-field equations of motion of a BEC in a tilted optical lattice, 
the DNLSE:
\begin{equation}
\label{1}
i\;\hbar\dot{a}_l=\epsilon_l a_l-\frac{J}{2}(a_{l+1}+a_{l-1}) 
+g|a_l|^2a_l \;,\quad \epsilon_l=dFl \,.
\end{equation}
Here $a_l(t)$ are the complex amplitudes of a mini BEC associated with the $l$\,th 
well of the optical potential, $J$ is the hopping or tunneling matrix element, $d$ 
the lattice period, $F$ the magnitude of the static force, and $g$ the nonlinear 
parameter, given by the product of the microscopic interaction constant $W$ and 
the filling factor $\bar{n}$ (the mean number of atoms per lattice site). It should 
be noted that Eq.~(\ref{1}) can also be derived as a tight-binding approximation for 
the discretized Gross-Pitaevskii equation. To simplify the equations we shall set the 
lattice period $d$ and the Planck constant $\,\hbar$ to unity in the following.
 
Throughout the paper we shall use the gauge transformation
\begin{equation}
a_l(t)\rightarrow \exp[-i(g+Fl)t] a_l(t)\,
\end{equation}
that eliminates the static term in Eq.~(\ref{1}). Note that the inclusion of $g$ in 
the transformation is optional and is done to facilitate the  
stability analysis below. The effect of the static force then appears as periodic 
driving of the system with the Bloch frequency $F$:
\begin{equation}
\label{1a}
i\dot{a}_l=-\frac{J}{2}\left( e^{-iFt} a_{l+1} +e^{+iFt} a_{l-1} \right)
+g\left(|a_l|^2-1\right)a_l  \;.
\end{equation}
An advantage of the gauge transformation is that one can impose periodic 
boundary conditions, $a_0(t)\equiv a_L(t)$, where we restrict ourselves to 
odd values of $L$. Equation (\ref{1a}) also appears as a canonical equation of 
motion generated by the Hamiltonian function
\begin{equation}
\label{1b}
H(t)=-\frac{J}{2}\sum_l\left(e^{iFt}a^*_{l+1}a_l+c.c.\right)
+\frac{g}{2}\sum_l|a_l|^2\big(|a_l|^2-2\big) \,.
\end{equation}

In this work, we shall be concerned mainly with almost uniform initial 
conditions $a_l(0)\approx 1$ which correspond to a BEC in the zeroth quasimomentum 
mode in the many-particle description in Sec.~\ref{sec3}.
Note that we do not normalize $\sum_l |a_l|^2$ to unity. Of course, this initial 
condition is an idealization of the real experimental  
situation, where initially only a finite number of wells are occupied.  
Nevertheless, an analysis of this situation provides useful estimates  
which, in fact, can also be applied to the case of non-uniform initial  
conditions \cite{09BOBECb}. 

It is convenient to switch to the Bloch-waves representation
\begin{equation}
b_k=L^{-1/2}\sum_{l=1}^L \exp(i\kappa l)\,a_l\ ,\quad k=0,\pm 1,\ldots ,\pm(L-1)/2\,, 
\end{equation}
where $\kappa=2\pi k/L$ is the quasimomentum ($-\pi\le\kappa<\pi$). After this 
canonical change of variables Eq.~(\ref{1a}) takes the form
\begin{eqnarray}
i\dot{b}_k&=&-J\cos(\kappa-Ft)b_k\nonumber \\[2mm]
&&+\frac{g}{L}\sum_{k_1,k_2,k_3} b_{k_1}b^*_{k_2}b_{k_3} 
\delta_{k,k_1+k_2-k_3}^{(L)} -gb_k\,,
\label{2}
\end{eqnarray}
where $\delta_{k,k'}^{(L)}$ is the Kronecker $\delta$ modulo  $L$. For 
strictly uniform initial conditions $a_l(0)=1$, Eq.~(\ref{2}) has the trivial solution
\begin{equation}
\label{3}
b_0(t)= \sqrt{L}\,\exp\Big(i\frac{J}{F}\,\sin(Ft)\Big) \;,\quad 
b_{k\ne0}(t)\equiv 0 \;,
\end{equation}
i.e.~a Bloch oscillation with period $T_B=2\pi/F$.
However, it is well known that for $g\ne 0$ the solution (\ref{3}) can be unstable 
with respect to a weak perturbation. The stability analysis of the solution 
(\ref{3}), resulting in the 
stability diagram in the parameter space of the system, was presented in 
Ref.~\cite{Kolo04a,Zhen04}. We extend this analysis below, mainly following 
Ref.~\cite{Kolo04a}, using methods of classical nonlinear dynamics.

\subsection{Stability analysis}
\label{sec2a}

From a formal point of view, the BO (\ref{3}) is a periodic trajectory in the 
multi-dimensional phase space spanned by the $a_j$ or the $b_k$, respectively. 
Using the standard approach, we linearize 
Eq.~(\ref{2}) around this periodic trajectory which leads to pairs of 
coupled equations,
\begin{eqnarray}
\nonumber
i{\dot b}_{+k}=-J\cos(\kappa-Ft)b_{+k}
+\frac{g}{L}|b_0|^2 b_{+k}+\frac{g}{L}b_0^2 b^*_{-k} \\
\label{4}
i{\dot b}_{-k}=-J\cos(\kappa+Ft)b_{-k}
+\frac{g}{L}|b_0|^2 b_{-k}+\frac{g}{L}b_0^2 b^*_{+k}
\end{eqnarray}
for $k\ne 0$, where the initial amplitudes $b_{\pm k}(0)$ are arbitrarily small. 
Substituting $b_0(t)$ from Eq.~(\ref{3}) and integrating Eq.~(\ref{4}) in time 
over $n$ Bloch periods, we get 
\begin{equation}
\label{5}
\left(\begin{array}{c} b_{+k}(t_n) \\ 
b^*_{-k}(t_n) \end{array}\right)=
\lambda_1^n \,\bf{b}_1+\lambda_2^n  \,\bf{b}_2 \;.
\end{equation}
Here $t_n=T_B n$ and $\lambda_{1,2}$ and $\bf{b}_{1,2}$ are the eigenvalues and 
eigenvectors of the stability matrix 
\begin{equation}
\label{6}
U^{(k)}=\widehat{\exp}\left[-ig\int_0^{T_B}\left(
\begin{array}{cc}
1&f(t)\\-f^*(t)&-1
\end{array}
\right)dt\right] \;,
\end{equation}
where the hat over the exponential function denotes time ordering, and
\begin{equation}
f(t)=\exp\big(i\,\frac{2J}{F}\,[1-\cos\kappa ]\sin(Ft)\,\big)\,. 
\end{equation}
Note that the determinant -- and therefore the product
of the eigenvalues -- of the (symplectic) stability matrix is equal to one. The 
trajectory is stable when both eigenvalues lie on the unit circle but becomes 
unstable when they merge on the real axis and go in and out of the unit circle.
In what follows we shall characterize this instability by the increment $\nu$, 
given by the logarithm of the modulus of the maximal eigenvalue: 
$\nu=\ln |\lambda_1|$, $|\lambda_1|\ge|\lambda_2|$. The 
increment of the dynamical instability is parameterized by the quasimomentum 
$\kappa$ and, hence, there are $(L-1)/2$ different increments $\nu^{(k)}\ge 0$. 
For a stable BO, all of them should vanish simultaneously. Further details 
concerning the stability analysis can be found in Appendix \ref{sec-stability}.
%
\begin{figure}[t!]
\center
\includegraphics[width=8.5cm, clip]{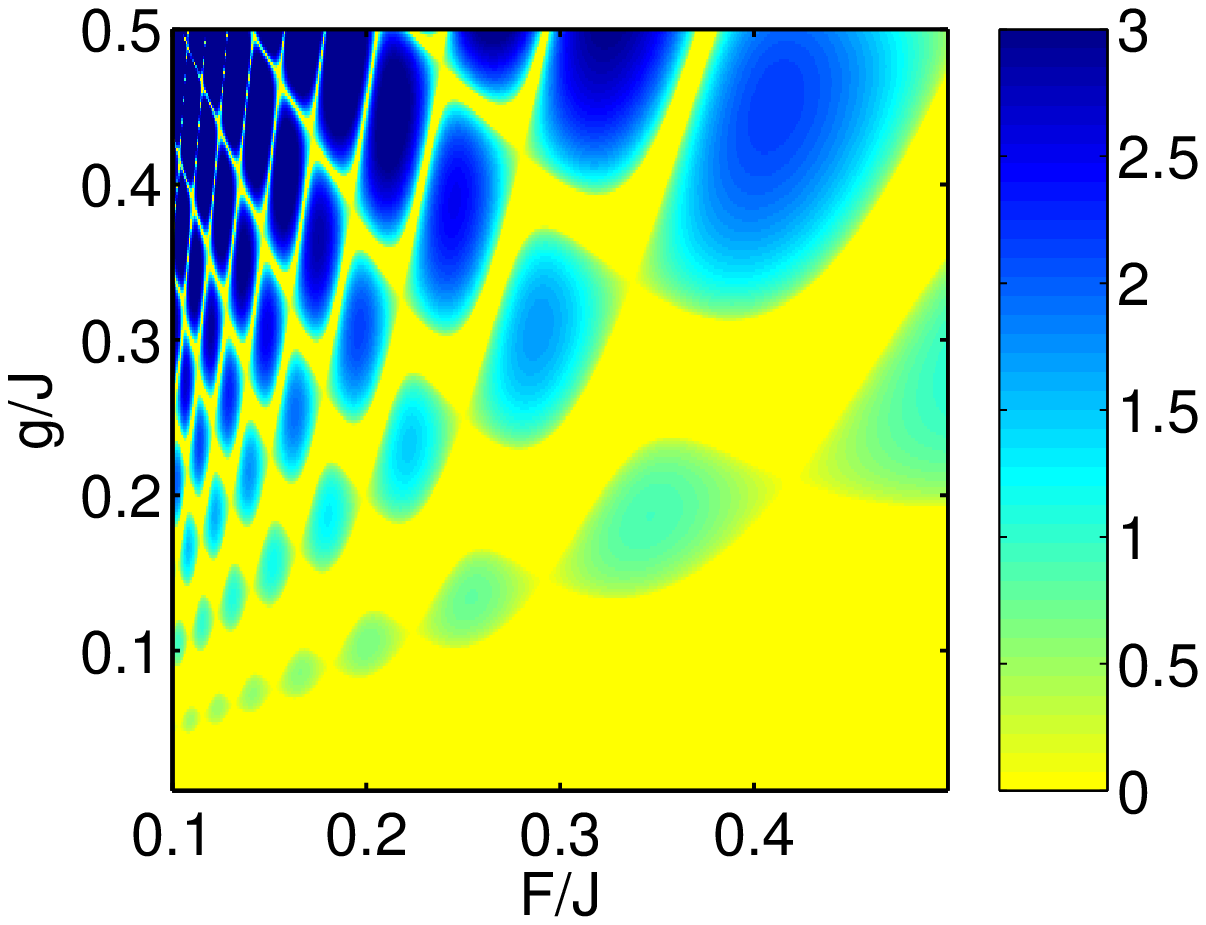} 
\includegraphics[width=8.5cm, clip]{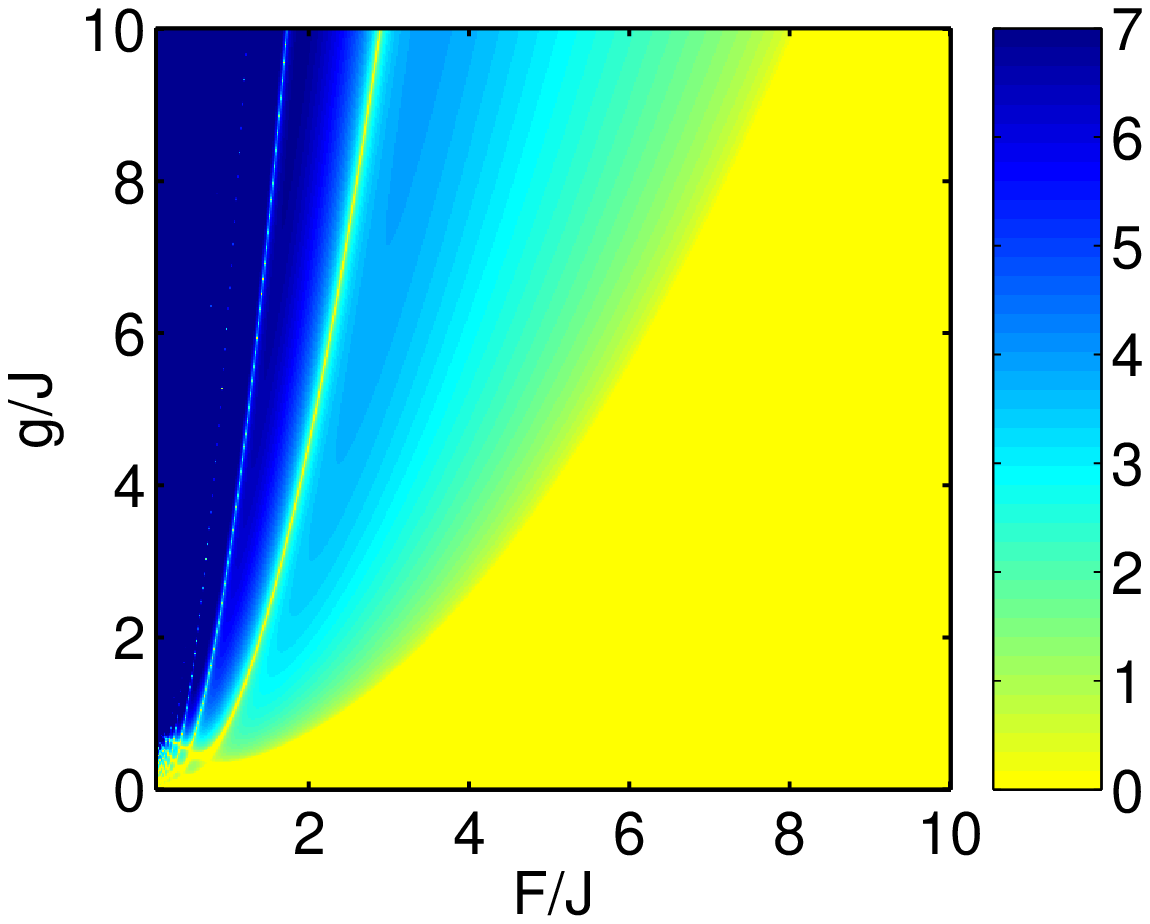} 
\caption{Increment of the dynamical (modulation) instability $\nu$ as a
function of the static force magnitude $F/J$ and the interaction $g/J$ for
$L=3$ sites. The yellow region in the lower panel corresponds to $\nu^{(k)}=0$ 
for all $k$, i.e.~the Bloch
oscillation is stable. A magnification of the lower left corner is depicted in 
the upper panel
and reveals a web of additional stability regions.}
\label{fig1a}
\end{figure}

\begin{figure}[t!]
\center
\includegraphics[width=8.5cm, clip]{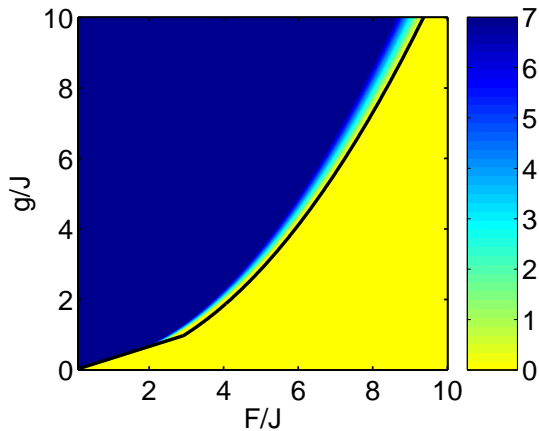} 
\caption{Same as Fig.~\ref{fig1a} (lower panel), however for $L=63$ sites. Also 
shown is the boundary given by Eq.~(\ref{7}) as a black curve.}
\label{fig1a63}
\end{figure}

As an example we show the increment of the dynamical instability for 
the system with only $L=3$ sites in dependence of $F/J$ and $g/J$ in 
Fig.~\ref{fig1a}. It is seen in the figure that the parameter space 
of the system is divided into two parts by a critical boundary. 
If the number of sites is increased, the instability regions grow and the boundary 
approaches approximately  the curve 
\begin{equation}
\label{7}
F_{cr}\approx\left\{
\begin{array}{ll}
3\,g  &,\quad F<2.9\,J \\[2mm] 
2.96\sqrt{gJ}\ &,\quad F>2.9\,J
\end{array}\right.
\end{equation}
(see \cite{Zhen04}). Figure \ref{fig1a63} shows the stability diagram for $L=63$ lattice sites
together with the boundary curve (\ref{7}).

\begin{figure}[t!]
\center
\includegraphics[width=8.5cm, clip]{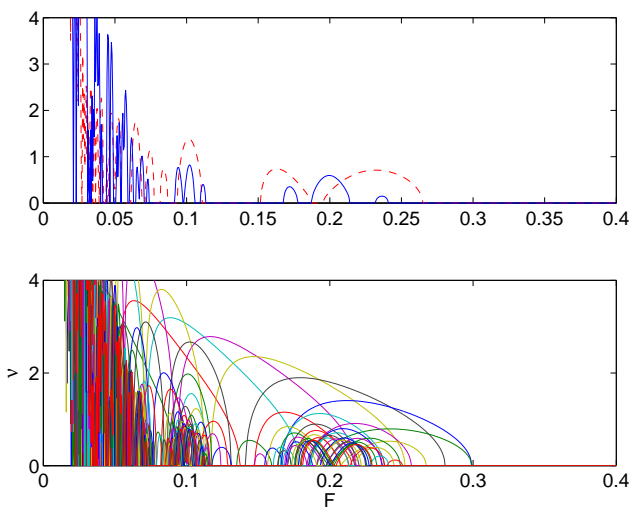}
\caption{Increments of the modulation instability $\nu^{(k)}=\nu^{(k)}(F)$, 
$k=0,\,1,\ldots$ for $J=1$,  $g=0.1$ and $L=5$ sites (upper panel) as well as 
$L=63$ sites (lower panel). With increase of the lattice size the instability 
regions for different $k$ overlap and cover the whole region left to the critical line.}
\label{fig2}
\end{figure}

In the right region of the diagrams $\nu=\nu(F,g)\equiv0$ holds
and BOs are always stable. Physically that means that a sufficiently large static 
force ensures stability even in the presence of nonlinearity. In the left hand side 
of the diagrams BOs are typically unstable but a closer inspection reveals a web of 
stability regions for smaller values of the parameters as can be seen in the upper 
panel in Fig.~\ref{fig1a}. It should be noted that these stability regions are a 
particular property of systems with a small number of lattice sites. 
If $L$ is increased, the regions where $\nu^{(k)}>0$ overlap (see Fig.~\ref{fig2}) 
and for any point in the left side of the diagram there is at least one strictly 
positive increment of the dynamical instability. 
\subsection{Relation to chaos}
\label{sec2b}
In this section we investigate the implications of the stability on the 
dynamics of a classical ensemble of trajectories. We will address the relation 
between dynamical instability, classical chaos and the so-called self-thermalization, 
that is, equipartition of the energy between the different quasimomentum modes. 
Related questions have been studied in \cite{Berm84} in the context of one of the 
standard systems of classical chaos, the Fermi-Pasta-Ulam system \cite{FPUchaos}. 
There it was pointed out that a positive increment $\nu$ of the dynamical instability is 
a necessary but not sufficient condition for the onset of developed chaos in a chain 
of coupled nonlinear oscillators. Here the term `developed chaos' characterizes a 
situation where a chaotic trajectory explores the whole energy shell. Developed chaos 
also implies equipartition of the energy between the eigenmodes of the chain, a 
phenomenon often referred to as thermalization. 

\begin{figure}[h!]
\center
\includegraphics[width=8.0cm, clip]{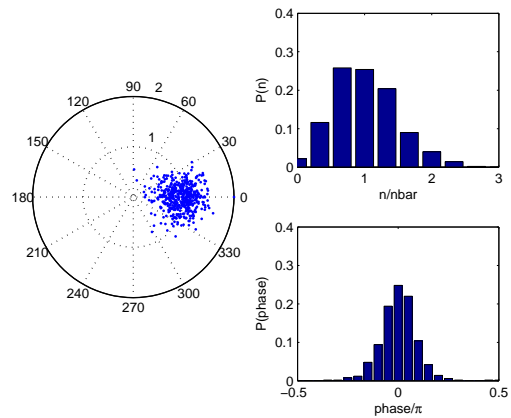}
\caption{Classical ensemble representing the many-particle initial state
for \,$N=15$, $L=5$ and 100 realizations. The characteristic width 
of the distribution  is proportional to $\bar{n}^{-1/2}$.}
\label{figens}
\end{figure}

The consideration of an ensemble of classical trajectories as compared to a single 
trajectory is well suited for understanding the systems behavior in the present 
context for two main reasons. First it is a convenient method to capture the 
generic features of perturbed initial conditions due to an out-averaging of the special 
behavior of an individual trajectory slightly differing from the BO (\ref{3}). The 
second reason is that the mean-field description is an approximation in the spirit of 
a classical limit of the many-particle system which has to be regarded as the more 
fundamental description. This many-particle system, however, cannot be associated 
with a point in the (classical) mean-field phase space, but is rather equipped with 
a finite width due to the uncertainty principle, where the width decreases with 
increasing particle number. Thus the natural counterpart of the many-particle system 
within the mean-field description is a phase space distribution rather than a single 
point. This phase space distribution can be conveniently replaced with a finite 
ensemble of classical trajectories for practical purposes. Further details of this 
considerations can be found in \cite{07phase} and Appendix \ref{sec-ensemble}. As an 
example we depict in Fig.~\ref{figens} the amplitudes $a_l$, $l=1,\ldots,L$ of an 
initial classical ensemble for the parameters $N=15$, $L=5$ considered in Sec.~\ref{sec3} 
for 100 realizations. The histograms show the corresponding probability 
distributions for the populations $|a_l|^2$ and the phases.
%

A convenient quantity for the characterization of the Bloch dynamics in the higher 
dimensional classical phase space is  the classical momentum given by 
\begin{equation}
p(t)=\frac{1}{2i}\Big[\sum_l a^*_{l+1}a_l \,e^{-iFt}-c.c.\Big]
= \sum_k |b_k|^2\sin(\kappa-Ft).
\label{momentum}
\end{equation}
For the periodic solution (\ref{3}), that is, the BO, the momentum oscillates 
in a cosine manner between $-1$ and $1$ with the Bloch frequency $F$. For neighboring 
initial values the behavior may differ considerably, depending on the stability of 
the BO: If the BO is unstable we expect an exponential growth of the initial deviation 
connected to classical chaos, leading to thermalization. On the other hand, the 
naive expectation in the stable region is that a small initial deviation leads to 
a small deviation in the overall trajectory and therefore averaging over an ensemble 
does not change the Bloch oscillation behavior in principle. However, we are going 
to argue that this is only true in a certain range of the parameter space and indeed 
one can distinguish three regimes of classical motion instead of only `stable' or 
`unstable' behavior. In fact, we find that for large values of the static force $F$ 
the ensemble average leads to a breakdown of the BOs and even a thermalization in the 
absence of classical chaos.

\begin{figure}[b!]
\center
\includegraphics[width=8.5cm, clip]{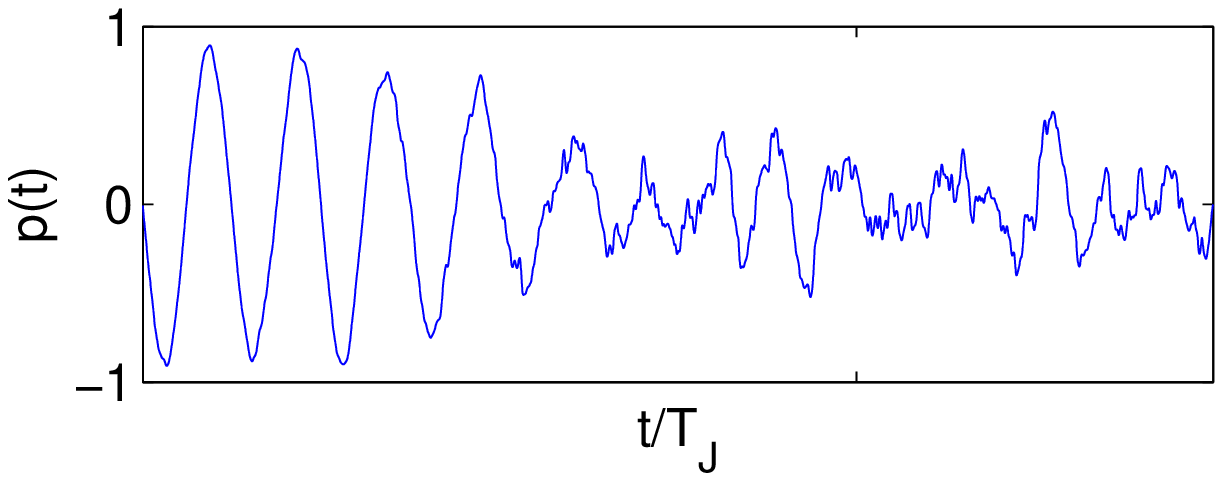}   
\includegraphics[width=8.5cm, clip]{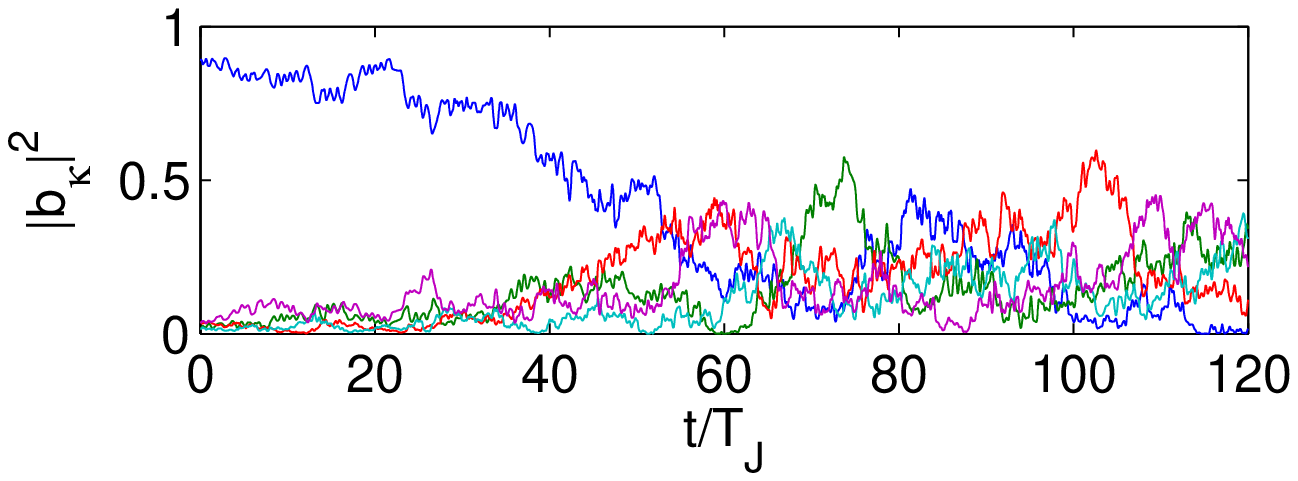}    
\caption{Mean-field BO of the momentum (upper panel) and the dynamics of the
population of quasimomentum modes (lower panel). These results for
a single trajectory oscillate erratically. Parameters are $L=5$, $J=1$, 
$g=0.1$, and $F=0.1$.}
\label{fig3F01a}
\end{figure}

Let us start our discussion with the unstable regime, that is, small values of $F$. 
The exponential growth of a deviation from the uniform initial condition is 
evidently related to chaos. In particular, as for the Fermi-Pasta-Ulam system 
\cite{FPUchaos}, for the system (\ref{1a}) positive increments of the 
dynamical instability are a necessary condition for the onset of developed chaos. 
To illustrate the impact on the Bloch dynamics we show the time evolution for a 
single trajectory of an ensemble mimicking an $N=15$ particle system in 
Fig.~\ref{fig3F01a}. The first panel in the figure depicts the momentum (\ref{momentum}) 
and the second panel shows the mode populations $|b_k(t)^2|$ of the quasimomentum 
modes $k$ as a function of time measured in units of $T_{\rm J}=2\pi/J$.

\begin{figure}[t!]
\center
\includegraphics[width=8.5cm, clip]{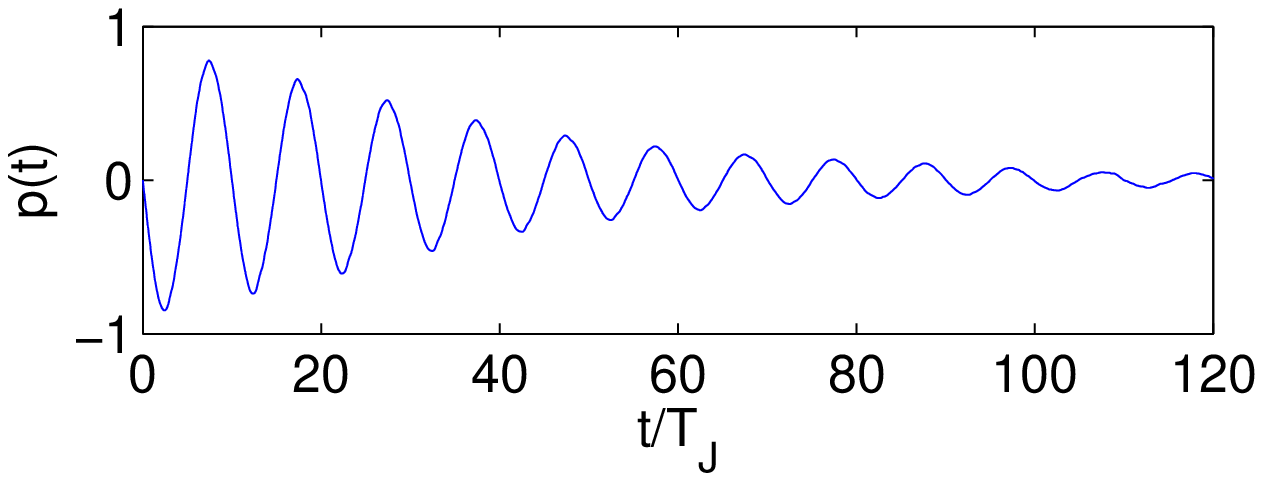} 
\includegraphics[width=8.5cm, clip]{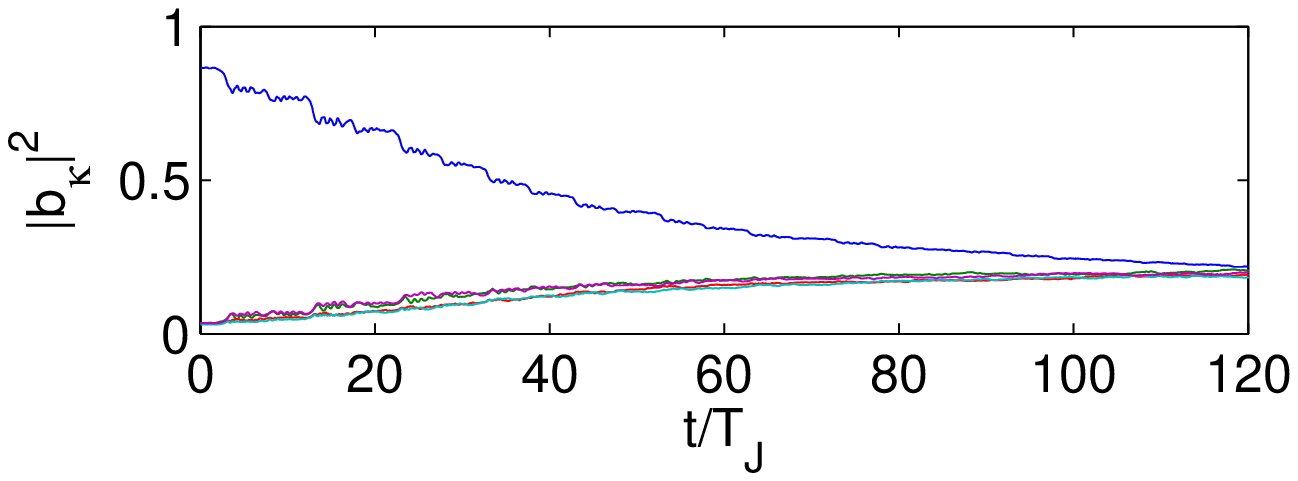}  
\caption{Mean-field BO of the momentum (upper panel) and dynamics of the
populations of quasimomentum modes, averaged over an ensemble of 1000 trajectories of
initial conditions (\ref{ensemble2}). The other parameters are the same 
as in Fig.~\ref{fig3F01a}.  An equipartition
between the quasimomentum modes results in the decay of BO.}
\label{fig3F01b}
\end{figure}

For a single run, the momentum $p(t)$ and the mode populations $|b_k(t)|^2$ 
start to oscillate irregularly after a transient time $t_\nu\sim\ln(\epsilon/\nu)$,  
$\epsilon\sim \bar{n}^{-1/2}$, required for the modes with $k\ne0$ to take 
non-negligible values. These erratic oscillations are smoothed by averaging over 
the ensemble (consisting of 1000 trajectories in the present example) as shown in 
Fig.~\ref{fig3F01b}. In the ensemble average one observes a damped BO of $p(t)$ 
and an equipartition of the mode populations converging to the values of $1/L$, 
i.e., a thermalization. It is clearly seen in the figure that the rate of 
thermalization actually determines the decay rate of the BO. 
In Sec.~\ref{sec3} we will demonstrate that this (classical) mean-field dynamics
agrees remarkably well with the full quantum many-particle  behavior
in view of the small number of three particles per site.

\begin{figure}[b!]
\center
\includegraphics[width=7.5cm, clip]{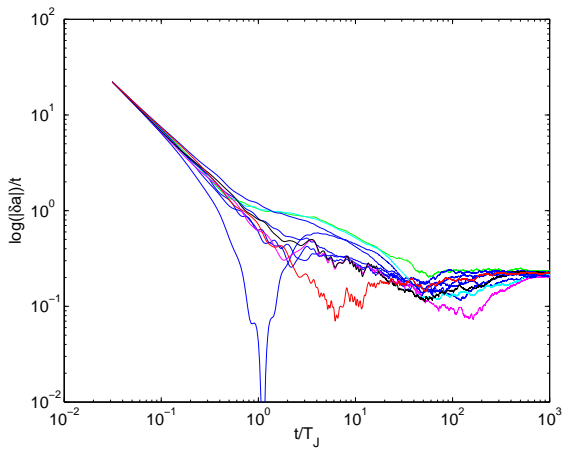}
\caption{Finite time Lyapunov exponent $\lambda(t)$ for ten different trajectories 
from the ensemble (\ref{ensemble2}). (Same parameters as in Fig.~\ref{fig3F01a}.)}
\label{fig3c}
\end{figure}
One gets further insight into the relation to chaos by studying the finite
time Lyapunov exponent 
\begin{equation}
\lambda(t)=\ln|\delta {\bf a}(t)|/t \;,
\end{equation}
where $\delta {\bf a}(t)$ evolves in the tangent space according to
the linear equation
\begin{equation}
i\frac{d \delta {\bf a}}{dt}=M[{\bf a}(t)] \,\delta {\bf a}(t)
\end{equation}
(see Appendix \ref{sec-stability}). As an example, Fig.~\ref{fig3c} shows the 
behavior of $\lambda(t)$ for ten different trajectories from an ensemble 
(\ref{ensemble2}). It is seen that $\lambda(t)$ converges to some constant value, 
so that the mean Lyapunov exponent $\langle\lambda\rangle$ is a well defined 
quantity. Since the unstable periodic trajectory analyzed
in subsection \ref{sec2a} is a member of this ensemble, the maximal
increment of the modulation instability $\nu={\rm max}_k \nu^{(k)}$ provides
a reliable estimate for $\langle\lambda\rangle$. Still, because the mean
Lyapunov exponent depends on the ensemble (i.e. on the value of the
filling factor $\bar{n}=N/L$), generally $\langle\lambda\rangle\ne\nu$. In
particular, $\langle\lambda\rangle$ is found to be a smooth function
of $F$, while the maximal increment of the modulation instability is a
non-analytic function of $F$ with discontinuous first derivative. As a
consequence, when we cross the critical line in the stability diagram,
the system shows a smooth transition from the regime of decaying BOs to
the regime of persistent BOs. (For example, for the parameters of
Fig.~\ref{fig3F01b} this change happens in the interval $0.1\lesssim F
\lesssim 0.4$.)

\begin{figure}[b!]
\center
\includegraphics[width=8.5cm, clip]{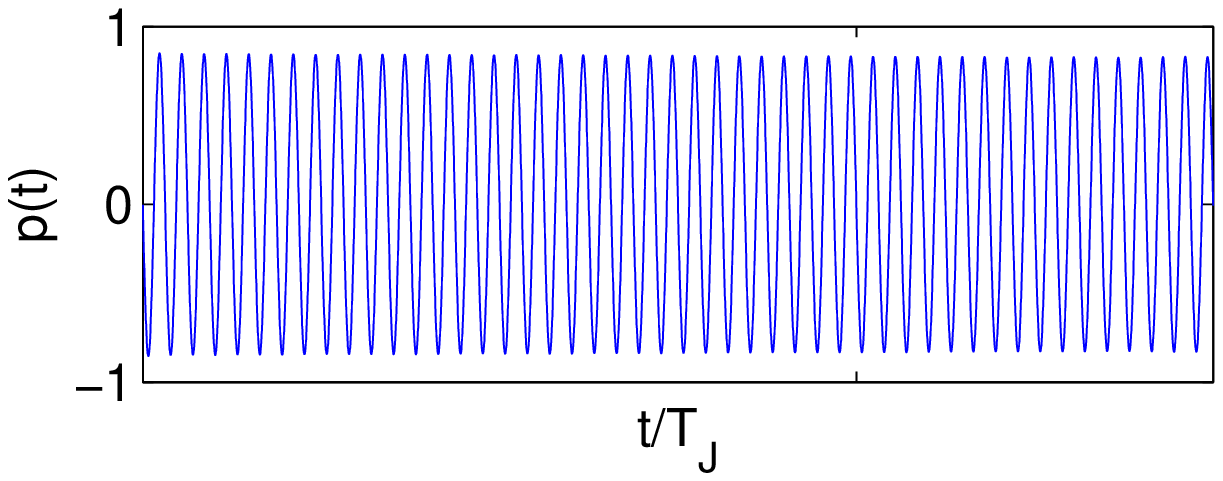}  
\includegraphics[width=8.5cm, clip]{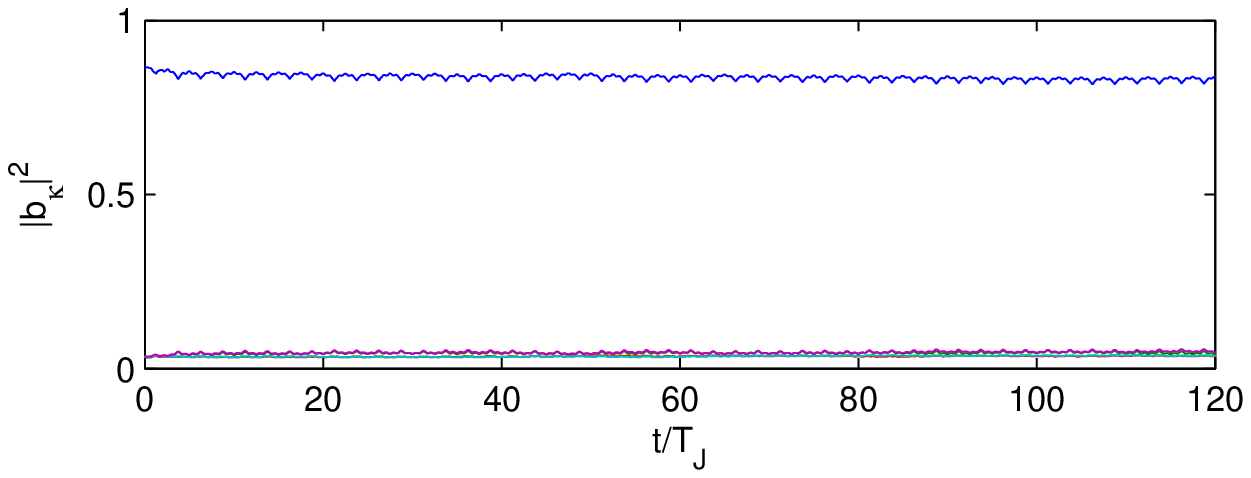}  
\caption{Same as Fig.~\ref{fig3F01b}, however for $F=0.4$ showing a stable
Bloch oscillation.}
\label{fig3F04}
\end{figure}

We now turn to the parameter regime of stable BOs, for larger values of $F$ where 
the increment is zero. Here we will distinguish two different types of
behavior in the 
ensemble average: Persistent BOs for intermediate values of $F$ and decaying BOs 
connected to thermalization in the limit of large $F$. The regime of persistent BO 
is shown in Fig.~\ref{fig3F04} for an example with $F=0.4$. No energy exchange 
between quasimomentum modes is seen which means that the system dynamics is at least 
locally regular. In other words, for the considered moderate $F$ the periodic 
trajectory (\ref{3}) is surrounded by a stability island. Moreover, the size of 
this stability island should be large enough as compared to the characteristic 
width of the distribution (\ref{ensemble2}), so that the majority of the 
trajectories are stable. In this case, the ensemble averaging is of little 
influence, resulting only in a small decrease of the amplitude.

\begin{figure}[t!]
\center
\includegraphics[width=8.5cm, clip]{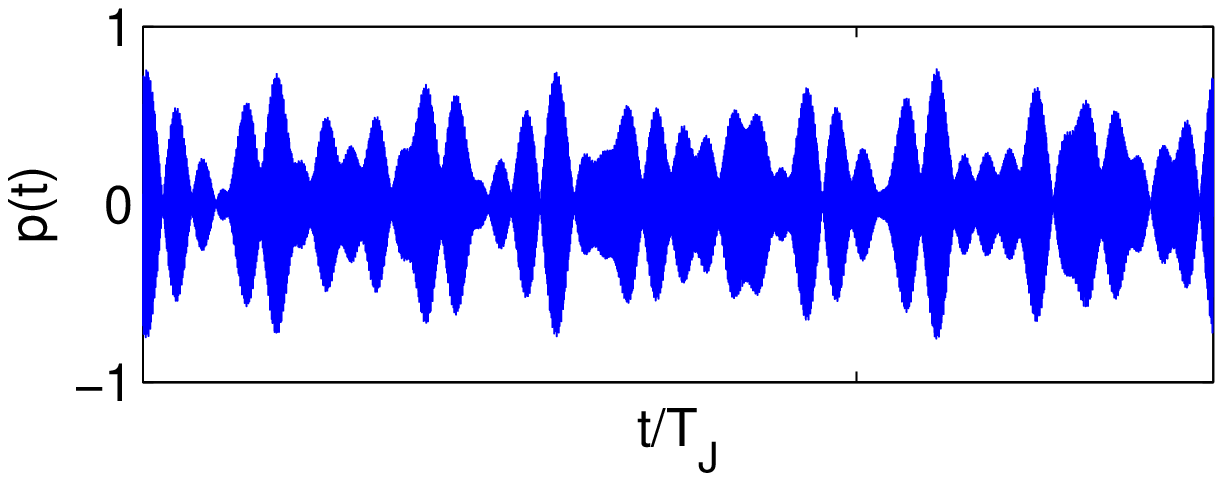} 
\includegraphics[width=8.5cm, clip]{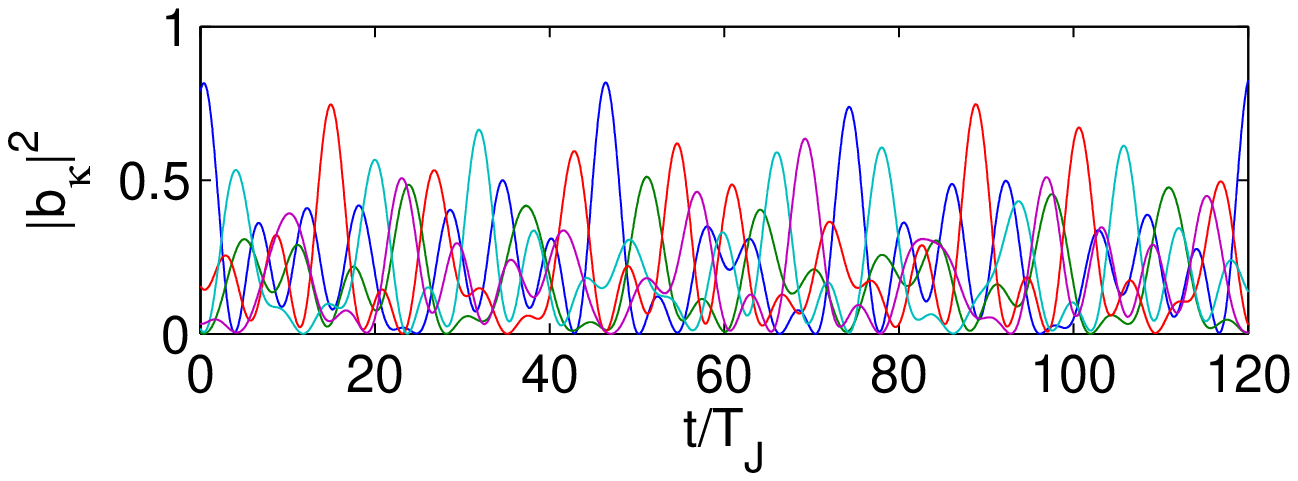} 
\caption{Same as Fig.~\ref{fig3F01a}, yet for $F=10$ in the
strong field region.}
\label{fig3F10a}
\end{figure}

\begin{figure}[b!]
\center
\includegraphics[width=8.5cm, clip]{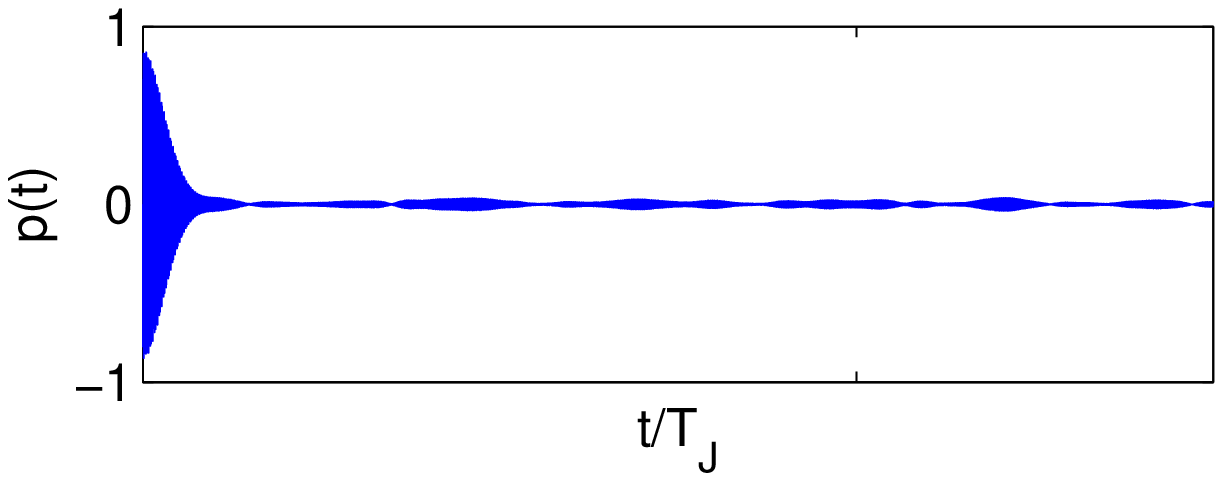} 
\includegraphics[width=8.5cm, clip]{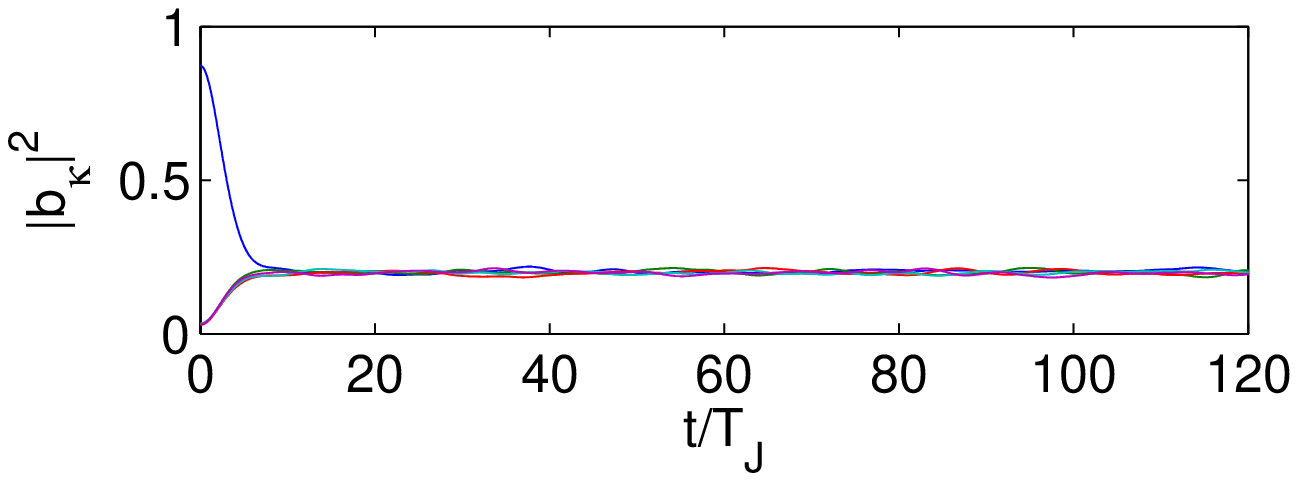} 
\caption{Same as Fig.~\ref{fig3F01b}, yet for $F=10$ in the
strong field region.}
\label{fig3F10b}
\end{figure}

The behavior for large values of $F$ and its relation to classical chaos is 
more surprising: An analysis of the phase-space structure of the system 
(\ref{1b}) reveals the volume of the regular component to grow with $F$ and 
for $F\rightarrow\infty$ the system becomes integrable \cite{Kolo04a}. 
(This integrable regime has already been observed in \cite{Kolo03}.)
Thus, 
one would first expect persisting BOs. However, the observed behavior is 
quite different. As an illustration we show an example for an individual 
trajectory in Fig.~\ref{fig3F10a}. Here one observes a quasiperiodic behavior. 
When averaged over an ensemble this leads to a decay of the BO as depicted in 
Fig.~\ref{fig3F10b}. This behavior can indeed be understood analytically. In 
the limit $F\to\infty$ the populations of the lattice sites are frozen and the 
amplitudes $a_l$ evolve only in the phase according to
\begin{equation}
\label{frozen}
a_l(t)\approx a_l(0)\exp\left[-ig(|a_l(0)|^2-1)\,t\right] \;.
\end{equation}
The evolution (\ref{frozen}) for the amplitudes $a_l(t)$ immediately implies 
the observed quasiperiodic dynamics of the amplitudes $b_k(t)$ (see 
Fig.~\ref{fig3F10a}). It can be shown that the decay due to the dephasing arising 
for the ensemble averaged dynamics obeys an  $\exp(-\gamma_r t^2)$-law, where the 
coefficient $\gamma_r$ is proportional to $g$ and inversely proportional to 
$\bar{n}$. Remarkably, the quasiperiodic dynamics also implies an equipartition 
between  the quasimomentum modes (see Fig.~\ref{fig3F10a}) in the absence of 
classical chaos. However, there are characteristic differences compared to the 
decay and the thermalization processes introduced by classical chaos: First, 
the relaxation constant $\gamma_r$ for the dephasing decay depends on the 
characteristic width of the distribution function (\ref{ensemble2}) and decreases 
as $1/\bar{n}$ in the semiclassical limit. On the contrary, the relaxation 
constant $\gamma_c$ for the chaotic decay decreases only as $1/\ln\bar{n}$. 
Second, the chaotic decay follows  
$\langle p(t) \rangle=\exp(-\gamma_c t)\sin(\omega_B t)$, while for the 
dephasing decay we have $\langle p(t) \rangle=\exp(-\gamma_r t^2)\sin(\omega_B t)$. 

Going ahead we note that both
the regimes of stable BO and the one of decaying BO in the presence 
of classical chaos closely resemble the full many-particle dynamics when averaged 
over an ensemble. However, in what follows we shall find that there is an additional 
many-particle feature in the dynamics in the third regime of large $F$. Here 
the decay of the Bloch oscillations is present in the many-particle system as 
well and for short times the mean-field ensemble even quantitatively resembles 
the many-particle dynamics. However, the many-particle system shows a periodic 
revival behavior as already pointed out in \cite{Kolo03} which, as a pure quantum 
phenomenon, cannot be captured by the mean-field ensemble.
  

\section{Many-particle dynamics}
\label{sec3}
In this section we study the many-particle counterpart of the 
mean-field system discussed in the preceding section. 
This is described by the driven Bose-Hubbard Hamiltonian
\begin{equation}
\label{20}
\widehat{H}(t)= 
-\frac{J}{2}\sum_l\Big(e^{iFt}\hat{a}^\dag_{l+1}\hat{a}_l+h.c.\Big)
+\frac{W}{2}\sum_l \hat{n}_l(\hat{n}_l-2) \;.
\end{equation}
We also confine the system to $L$ lattice sites ($L$ chosen to be odd) and
apply periodic boundary conditions. Then the Hilbert 
space of system (\ref{20}) is spanned by the Fock states 
$|{\bf n}\rangle=|n_1,n_2,\ldots,n_L\rangle$, where $\sum_l n_l=N$ is the total number 
of atoms. The dimension of the Hilbert space 
is equal to $\frac{(N+L-1)!}{N!(L-1)!}$\,. Using the canonical 
transformation $\hat{b}_{k}=L^{-1/2}\sum_l\exp(i2\pi kl/L)\,\hat{a}_l$, the 
Hamiltonian (\ref{20}) can be presented in the form
\begin{eqnarray}
\label{21} 
\widehat{H}&=&-J\sum_{k} \cos(\kappa-Ft)\,\hat{b}^\dag_{k}\hat{b}_{k}\nonumber\\
&&\ \ +\frac{W}{2L}\sum_{k_i}\hat{b}^\dag_{k_1}\hat{b}_{k_2}
\hat{b}^\dag_{k_3}\hat{b}_{k_4}\,\delta_{k_1+k_3,k_2+k_4} \;,
\end{eqnarray}
where we omit the constant term $W\sum_k \hat{b}_k^\dagger \hat{b}_k$.
The basis vectors of the Hilbert space are now given by the Fock 
states in the quasimomentum representation:
$|n_{-(L-1)/2},\ldots,n_{-1},n_0,n_{+1},\ldots,n_{(L-1)/2}\rangle$\,.
In the coordinate representation the Fock and the quasimomentum Fock states 
are given by the symmetrized product of the Wannier and Bloch functions, 
respectively. Here we are interested in the solution of the time-dependent 
Schr\"odinger equation with the Hamiltonian (\ref{21}) for initial conditions 
given by a BEC of atoms in the zero quasimomentum state, i.e. 
$|\Psi_0\rangle=|\ldots,0,N,0,\ldots\rangle_{q}$. This is, in fact, equivalent 
to an $SU(L)$ coherent state, namely
\begin{equation}
\label{31} 
|\Psi_0\rangle=\frac{1}{\sqrt{N!}}\Big(\frac{1}{\sqrt{L}}\sum_{l=1}^L
\hat a^\dagger_l\Big)^N\!|0\rangle=\sum_{\bf n} c_{\bf n} 
|{\bf n}\rangle
\end{equation}
where $|{\bf n}\rangle$ with ${\bf n}=(n_1,\ldots,n_L)$ 
is a Fock state
and $c_{\bf n}=\sqrt{\frac{N!}{L^Nn_1!\cdots n_L!}}$. 
In general the $SU(L)$ coherent states are equivalent to the fully condensed 
states, where our special choice approximately corresponds to the ground 
state of the system at $F=0$ if the condition $W\ll J$ is satisfied.

\subsection{Floquet-Bogoliubov states}
\label{sec3a}

First we address the question of a manifestation of the dynamical instability 
in the many-particle quantum system. It is argued below that, similar to the static 
case $F = 0$, the quantum  
counterpart of the dynamical instability is Bogoliubov's depletion of the condensate
\cite{Legg01}. We begin with an alternative derivation of the common 
Bogoliubov spectrum 
for $F=0$  \cite{Kolo0607a} which we shall then adopt to the case $F\ne0$.

For an infinite number of particles, the Bogoliubov states can be constructed 
from $|\Psi_0\rangle$ by applying the depletion operators 
\begin{eqnarray}
\label{22b} 
\widehat{D}^{(k)}=\sum_{n=0}^\infty c_n^{(k)} 
\left(\hat{b}^\dag_{-k}\hat{b}^\dag_{+k}\hat{b}_0\hat{b}_0\right)^n \,,
\end{eqnarray}
which transfer particles from the zero quasimomentum state to the
states $\pm k$\,:
\begin{eqnarray}
\label{22a} 
|\Psi\rangle=\prod_{k>0}\widehat{D}^{(k)}|\Psi_0\rangle \;.
\end{eqnarray}
In Eq.~(\ref{22b}) the coefficients $c_n^{(k)}$ should be determined 
self-consistently so that the wave function $|\Psi\rangle$ satisfies the stationary 
Schr\"odinger equation with the Hamiltonian (\ref{21}) with $F=0$,
\begin{equation}
\label{23} 
\widehat{H}(F=0)|\Psi\rangle=E |\Psi\rangle \;.
\end{equation}
Equations (\ref{22b}), (\ref{22a}) are equivalent to the ansatz
\begin{equation}
\label{24} 
|\Psi^{(k)}\rangle=\sum_{n=0}^\infty c^{(k)}_n 
|..,n,..,N-2n,..,n,..\rangle_{q}  \;,
\end{equation}
where $n$ particles are redistributed from the zero quasimomentum state
to the state $k$ and to the state $-k$. Note that here,
as well as in (\ref{22b}), (\ref{22a}), the limit 
$N\rightarrow\infty$, $W\rightarrow0$ with constant $g=WN/L$ is assumed,
which justifies a factorization of the eigenvalue problem (\ref{23}) into
$(L-1)/2$ independent eigenvalue problems.
Substituting the ansatz (\ref{24}) into (\ref{23}) and taking the above 
limit the original problem reduces to the diagonalization of a 
tridiagonal hermitian matrix $A^{(k)}$ with matrix elements
\begin{equation}
\label{25}
A^{(k)}_{n,n}=2(g+\delta)n \quad  ,\quad A^{(k)}_{n,n+1}=g(n+1)
\end{equation}
where $\delta=J\left(1-\cos\kappa \right)$. The spectrum of the matrix $A^{(k)}$ is 
equidistant with a level spacing given by the Bogoliubov frequency 
$\Omega^{(k)}=2\sqrt{2g\delta+\delta^2}$. Correspondingly, the 
eigenvectors ${\bf c}^{(k)}$  of $A^{(k)}$ define the Bogoliubov states of the 
Bose-Hubbard model. It is worthwhile emphasizing that for a finite $N$ these 
Bogoliubov states provide only an approximation to the actual eigenstates of 
the Bose-Hubbard system. While this approximation can 
be quite accurate for the ground and low-energy states, it fails for the 
high-energy states \cite{Kolo0607a}.  
The BEC depletion of these states, defined by
\begin{equation}
\label{30} 
N_D=\sum_{k>0}\Big(\sum_{n=1}^\infty 2n\,|c^{(k)}_n|^2\Big) \;,
\end{equation}
provides a quantitative criterion for the validity of the Bogoliubov approach. 
Namely, if $N_D<<N$ then the approach is justified. On the contrary, 
$N_D \sim N$ (which is  
the case for high-energy states) means that the actual structure of the  
eigenstates has nothing to do with the presumed Bogoliubov structure  
(\ref{22a}). Furthermore, it is convenient to order the states
according to the depletion (\ref{30}).

Now we are in the position to discuss the case $F\ne0$. Since we are interested 
in the Bloch 
dynamics, it is useful to introduce the evolution operator over one Bloch period 
\cite{04bloch},
\begin{equation}
\label{26} 
\widehat{U}=\widehat{\exp}\Big[-i\int_0^{T_B} \widehat{H}(t) dt\Big] \;, 
\end{equation}
where $\widehat{H}(t)$ is given in (\ref{21}) and the hat over the exponential
function again
denotes time ordering. We are looking for those eigenstates of $\widehat{U}$,
\begin{equation}
\label{27} 
\widehat{U}\,|\Psi\rangle=\exp(-i2\pi E/F)\, |\Psi\rangle \,,
\end{equation}
which have the Bogoliubov structure (\ref{22a}). Substituting the ansatz (\ref{24}) into 
(\ref{27}) we obtain an equation for the coefficients $c_n^{(k)}$:
\begin{equation}
\label{28} 
\widehat{\exp}\,\Big[-i\int_0^{T_B}\!A^{(k)}(t)\, dt\Big] \,{\bf c}^{(k)}
= \exp\Big(-i\frac{2\pi E}{F}\Big)\, \bf{c}^{(k)}\;, 
\end{equation}
where the diagonal elements of the matrix $A^{(k)}(t)$ are now given by
\begin{equation}
\label{29}
A^{(k)}_{n,n}(t)=2[\,g+\delta\cos(Ft)]n \;,\quad \delta=J(1-\cos \kappa) \;.
\end{equation}
%
\begin{figure}[t!]
\center
\includegraphics[width=8.5cm, clip]{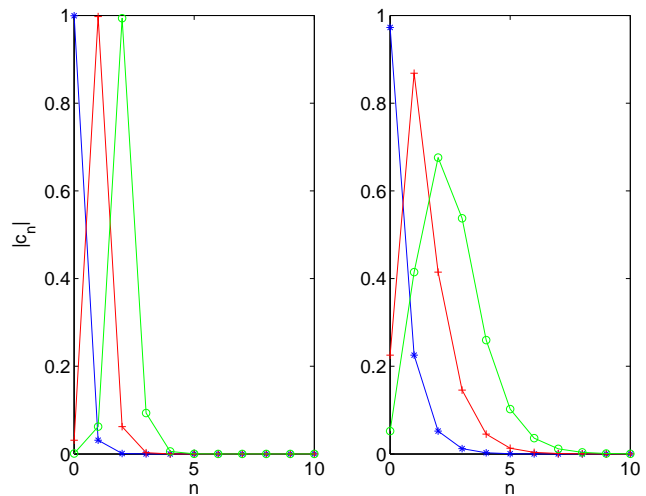}
\caption{Left panel: Coefficients of the first three Bogoliubov states 
$|\Psi\rangle=\sum_n c_n |n,N-2n,n\rangle$ of the three-sites Bose-Hubbard model. 
Right panel: Floquet-Bogoliubov states. (Parameters $J=1$, $g=0.1$ and $F=0.4$.)}
\label{fig4}
\end{figure}

As an illustration the right panel in Fig.~\ref{fig4} shows the coefficients of
the first three Floquet-Bogoliubov states in the quasimomentum Fock basis
for $L=3$, $g=0.1$, and $F=0.4$ \cite{remarkmf1}, where the depletion is 
$N_D=0.113$, $2.341$, 
and $4.568$ particles, respectively. For the sake of comparison, the left panel 
in 
the figure shows the Bogoliubov states for the same value of $g=0.1$, where the 
depletion is 
$N_D=0.002$, $2.006$, and $4.010$. It is seen in Fig.~\ref{fig4} that 
the depletion of the driven BEC is larger than the depletion of a stationary 
BEC which was found to be a typical situation in further numerical studies. 

\begin{figure}[t!]
\center
\includegraphics[width=8.5cm, clip]{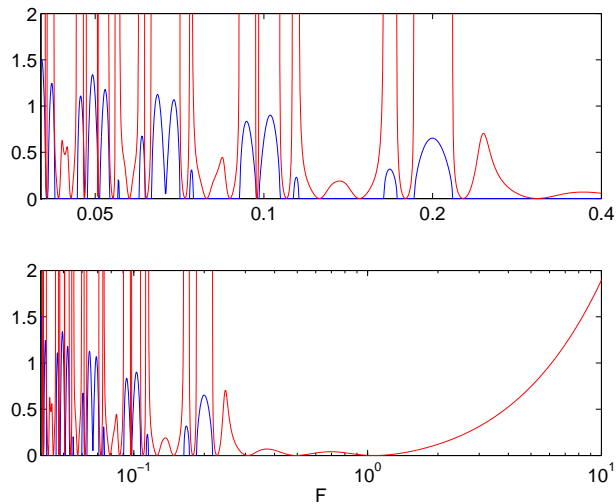}
\caption{Number of the depleted particles $N_D$ (red curve) as a function of the
field strength $F$ (logarithmic scale) compared to the classical 
increment of dynamical instability $\nu=\nu(F)$ (blue curve). Parameters are 
$J=1$, $g=0.1$, 
$L=3$. The upper panel magnifies the region of small $F$ in the lower panel.}
\label{fig5}
\end{figure}

\begin{figure}[t!]
\center
\includegraphics[width=8.5cm, clip]{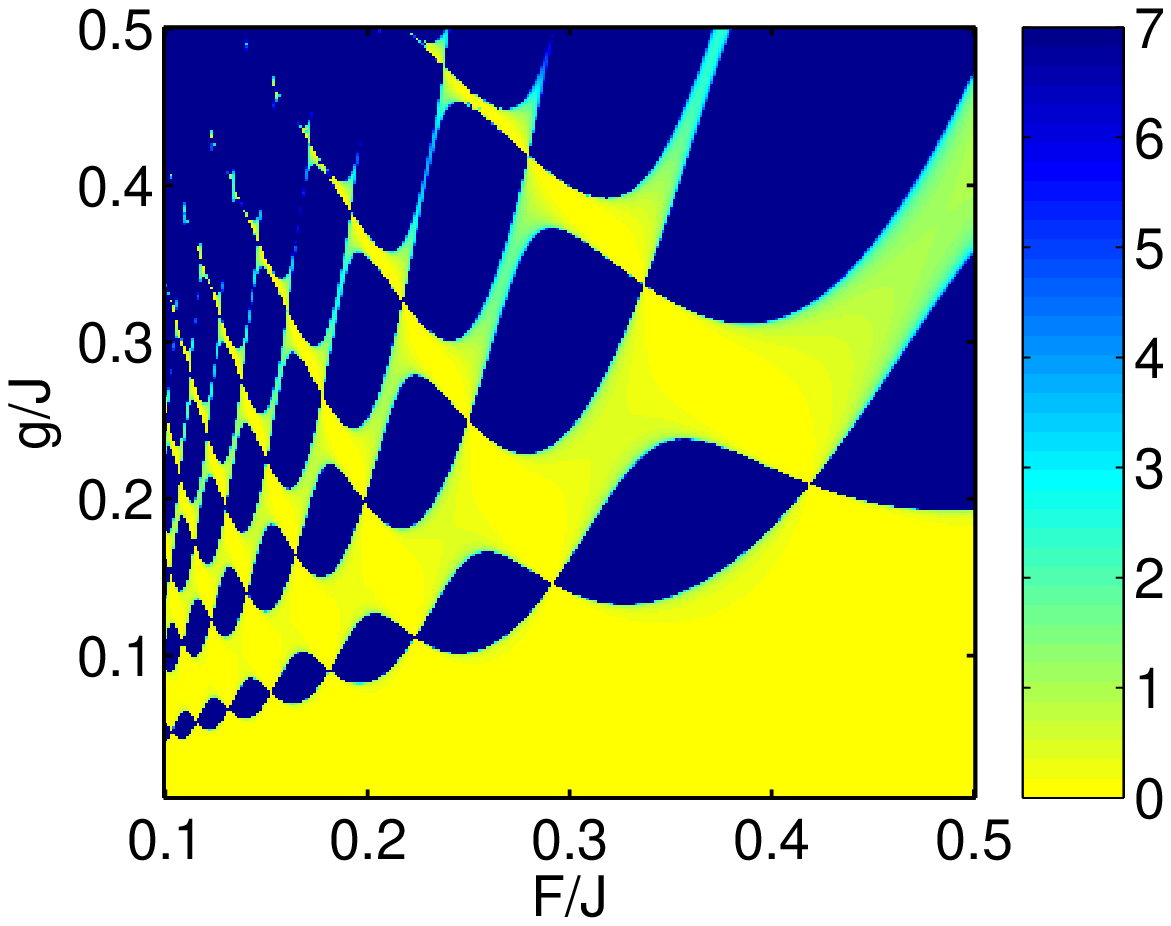} 
\includegraphics[width=8.5cm, clip]{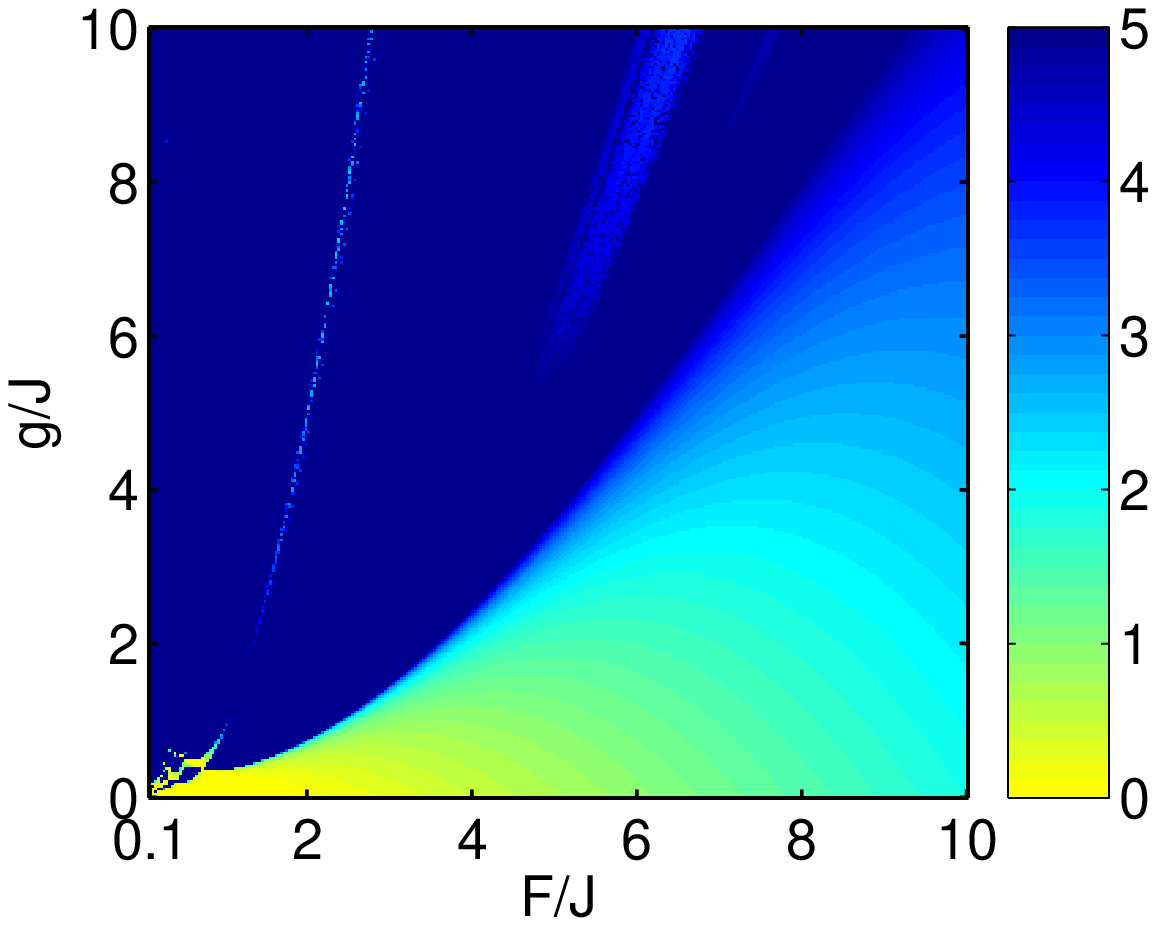} 
\caption{Number of the depleted particles $N_D$ as a
function of the static force magnitude $F/J$ and the interaction $g/J$ for
$L=3$ sites. The depletion is very small in the yellow
region of the lower panel. The upper panel magnifies the lower left corner of the 
lower panel and reveals additional regions of small depletion.}
\label{figDa}
\end{figure}

The depletion (\ref{30}) provides useful information on the BEC stability.
For the `ground' Floquet-Bogoliubov state the dependence $N_D=N_D(F)$ is depicted in 
Figs.~\ref{fig5} and \ref{figDa}.  
It is seen in Fig.~\ref{fig5} that, as a function of $F$, $N_D$ diverges 
at the points where 
the classical increment of the dynamical instability $\nu$ (also shown in the
figure) takes positive values. 
(More precisely, the depletion cannot be larger than $N$. However, since the
Bogoliubov theory refers to $N=\infty$, it can be formally considered as infinite.)
This `divergence' of $N_D$ means that in the regions of instability the 
eigenfunctions of the evolution operator differ considerably from the Bogoliubov 
structure. In fact, they are chaotic in the sense of quantum chaos \cite{04bloch}. 
It is also seen 
in the figure that the depletion increases linearly for $F\rightarrow\infty$. 
Thus the discussed Floquet-Bogoliubov states {\em cannot} be eigenstates of the 
evolution operator in the limit of large $F$, although the classical increments 
$\nu^{(k)}$ vanish identically for $F \rightarrow \infty$. We come back to this 
point in the next subsection.

The intervals of small depletion observed in Fig.~\ref{fig5} depend, of course,
also on the interaction $g$. The full parameter dependence of the depletion (\ref{30})
is shown in Fig.~\ref{figDa}. For relatively low values of $g/J$ and $F/J$ 
(upper panel) we find a highly organized web of stability regions, whereas the
behavior for larger parameters (lower panel) shows a simpler structure. 
These
many-particle results can be directly compared with the mean-field 
stability diagrams in Fig.~\ref{fig1a} which confirms
the relationship between BEC depletion and mean-field stability.

\begin{figure}[t!]
\center
\includegraphics[width=8.5cm, clip]{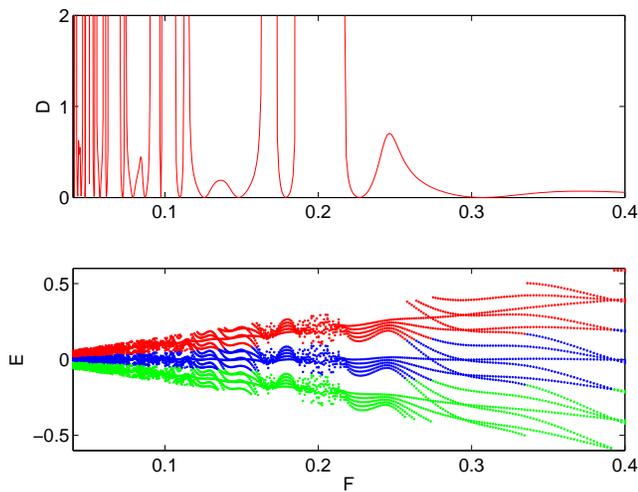}
\caption{Lower panel: Quasienergy spectrum of the Floquet-Bogoliubov states as a 
function of the field strength $F$ (same parameters as in Fig.~\protect\ref{fig5}).
Upper panel: Number of the depleted particles $N_D$.}
\label{fig6}
\end{figure}

For the sake of completeness we also briefly discuss the  
corresponding quasienergies $E$ (see Eq.~(\ref{27})) which are defined 
modulo $\,\hbar\,$ times the Bloch frequency, i.e. 
$E_{\alpha,j}=E_{\alpha,0}+jF$, $j =0,\,\pm 1,\ldots$\,.
The lower panel of Fig.~\ref{fig6} displays the quasienergies
of the five Floquet-Bogoliubov states with smallest $N_D$. 
The system parameters are $J=1$, $g=0.1$ and $L=3$ as in Fig.~\ref{fig5}.
For clarity, three Floquet zones are shown in the figure. 
As expected, the spectrum is  equidistant with a level spacing  
correlated to the depletion. Note the apparent irregularity in
the windows around $F=0.2$ and $F=0.17$, which is related to the region of
dynamical instability of the mean-field system. For comparison, the 
upper panel shows the number of depleted particles $N_D$ as in
Fig.~\ref{fig5} with a linearly scaled $F$-axis. 
In the regions of finite depletion the
most stable quasienergy states appear to be very sensitive against
a variation of $F$. 
 
\subsection{Bloch oscillations}
\label{3b}
The microscopic dynamics of BOs was considered earlier in a number of
papers summarized in Ref.~\cite{04bloch} with focus on the regime of low filling 
factors $\bar{n}=N/L\le1$. In the present work, to make a link to the mean-field 
dynamics, we simulate BOs for a relatively large filling factor.

We investigate the dynamics in dependence on the force $F$ for 
$N=15$ and $N=20$ atoms in a lattice with $L=5$ sites. Since the  
plots are very similar, only the results with $N=15$ are shown. 
The value of the microscopic interaction constant is set to 
$W=0.1/\bar{n}$, so that the macroscopic interaction constant is
$g=0.1$, and the hopping matrix element $J=1$. 
The initial wave function is chosen as the $SU(L)$ 
coherent state (\ref{31}). 

\begin{figure}[t!]
\center
\includegraphics[width=8.5cm, clip]{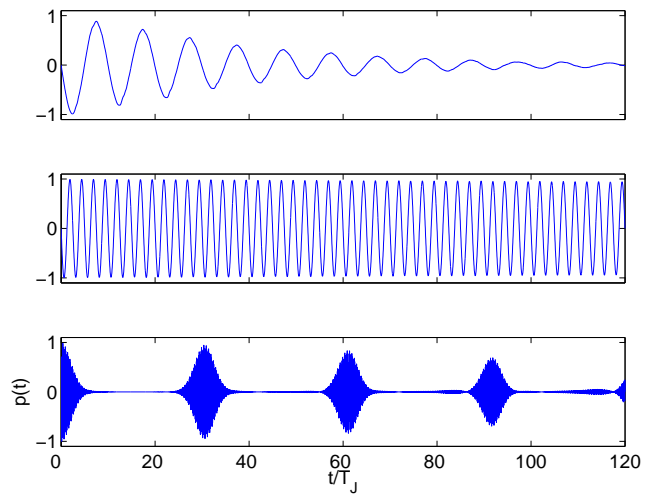}
\caption{Bloch oscillations of $N=15$ atoms in a lattice with $L=5$ sites 
for $J=1$, $W=0.1/3$, and $F=0.1$ (top), $F=0.4$ (middle), and $F=10$ (bottom).
Shown is the many-particle mean momentum $p(t)$ given in (\ref{Qmeanmom}).}
\label{fig7}
\end{figure}

In Fig.~\ref{fig7} we show the dynamics of the many-particle counterpart of the 
mean-field quantity (\ref{momentum}), that is, the expectation value of the 
many-particle momentum 
\begin{equation}
p(t)=\frac{1}{2iN}\,\Big\langle \psi(t)\Big|
\,\sum_l \hat a^\dagger_{l+1}\hat a_l \,e^{-iFt}-h.c.\Big]
\Big|\psi(t)\Big\rangle
\label{Qmeanmom}
\end{equation}
for $N=15$ and the three values of $F$ chosen in different dynamical regimes as 
shown in  Figs.~\ref{fig3F01b}, \ref{fig3F04} and \ref{fig3F10b}.

The upper panel of Fig.~\ref{fig7} corresponds to $F=0.1$, which falls into the 
region of dynamical instability of the mean-field system. Here
the quantum many-particle BO decays in very good agreement with
the mean-field ensemble shown in the upper panel of Fig.~\ref{fig3F01b}
(note that finer details are also reproduced). This demonstrates that the
ensemble averaged mean-field dynamics is capable of describing important
aspects of the full many-particle system \cite{07phase}.\\[1mm]
In the middle panel of Fig.~\ref{fig7} we have $F=0.4$, where the system is stable.
Here the quantum many-particle BOs persist in time, also agreeing
with the ensemble-averaged mean-field dynamics. As discussed above,
in this regime the ensemble average 
is of little influence, i.e.~the state is fully condensed and can be described by a 
single mean-field trajectory instead.

The third case, $F=10$, depicted in the lower panel of Fig.~\ref{fig7}, requires a 
separate consideration. Indeed, as mentioned in the previous subsection, in 
the limit of large $F$ the Floquet-Bogoliubov states are not eigenstates 
of the evolution operator (\ref{26}). Instead it can be shown that those are the 
Fock states $|{\bf n}\rangle$ \cite{04bloch}. Thus the time evolution of the wave 
function is given by
\begin{equation}
\label{32} 
|\Psi(t)\rangle=\sum_{\bf n} c_{\bf n} 
\exp\Big(-i\frac{Wt}{2}\sum_{l=1}^L n_l(n_l-1)\Big) |{\bf n}\rangle \;.
\end{equation}
Equation (\ref{32}) implies a periodic recovering of the initial state 
(\ref{31}) at times which are multiples of $T_W=2\pi/W$ and, hence, periodic 
revivals of BOs \cite{Kolo03}. 
It should be stressed that these revivals are a pure quantum many-particle
effect due to the finiteness of $W$ and $\bar{n}$. 
This constructive interference
cannot be explained within the ensemble-averaged 
mean-field approach. However, the breakdown can be described by the 
ensemble averaging and is due to dephasing, as explained in
Sec.~\ref{sec2b}.

\section{Summary}
\label{sec-concl}
We have studied an $N$-particle system, a Bose-Hubbard Hamiltonian
with linearly increasing on-site energies. This system can be conveniently
reduced to a finite lattice with $L$ sites by using gauge 
transformation and imposing periodic boundary conditions.

Such a model can be used to describe important features of realistic systems,
as for instance the dynamics of cold atoms or BECs in an optical lattice
under the influence of the gravitational field \cite{Daha96,Raiz97,Ande98,Mors01,
Cris02,Jona03,Ott04,04bloch_bec,Wimb05,Sias07,Gust08a,Fatt08},
or many-particle systems in ring shaped optical lattices as proposed in
\cite{Amic05} with additional driving.
It should be noted, however, that
it neglects a number of features. The space dimension is
reduced to a quasi one-dimensional setting, a decay of the system via
Zener transitions to higher bands is excluded,
the lattice is discretized and truncated, and the interaction is simplified.
Nevertheless, this model has been found to describe experimental
results quite well.
Further, the Bose-Hubbard model is of interest in its own right, as 
evident from the large number of studies exploring its properties which are 
remarkably rich.

In the present paper we have studied the Bose-Hubbard model for 
relatively small number of lattice sites $L\le 5$ but  
relatively large number of particles up to $N=20$ to make a  
link with the mean-field dynamics. Our aim was twofold: First, we have 
demonstrated how the modulation instability observed in
the mean-field system are manifested in the many-particle case.
We have shown that a reasonable measure of the $N$-particle
instability is provided by the quantity (\ref{30}). Second, we have explored the 
possibility to describe the time-evolution
of many-particle expectation values in terms
of mean-field trajectories, averaged over an ensemble constructed from
the $SU(L)$-phase space distribution of the initial many-particle state.
We found that this (averaged) mean-field dynamics agrees remarkably 
well with the full quantum many-particle behavior in a number of cases.
The only exception was the presence of many-particle revivals for strong 
fields which is a pure quantum phenomenon. 

These observations suggest an application of the mean-field ensemble
method to investigate the properties for larger lattices and larger
particle numbers where many-particle computations are much more 
difficult or virtually impossible. Furthermore, it will be of interest
to study the interrelation between classical chaotic motion
and stable or decaying Bloch oscillations for more lattice sites
where one can possibly make contact with recent related studies of the
force free, $F=0$, case, both for the mean-field \cite{Vill00,Cass08}
and the many-particle description \cite{Kolo0607a,Berm04}.

\vspace*{4mm}
\section*{Acknowledgments}
We thank D. Witthaut and F. Trimborn for valuable comments.
Support from the Deutsche Forschungsgemeinschaft
via the Graduiertenkolleg  ``Nichtlineare Optik und Ultrakurzzeitphysik''
is gratefully acknowledged. 

\appendix
\section{Stability Matrix}
\label{sec-stability}
It is instructive to consider the DNLSE (\ref{1a}) from the viewpoint of the general 
theory of nonlinear dynamics. Then the solution with the initial condition $a_l(t=0)=1$,
\begin{equation}
\label{40}
a_l(t)=\exp\Big(i\frac{J}{F}\sin(Ft)\Big) \,
\end{equation}
is nothing other than a periodic trajectory in a
$2L$-dimensional phase space. Thus one may address the question of 
stability of this periodic trajectory.

Denoting by 
$\delta {\bf a}=(\delta a_1,\ldots,\delta a_L,\delta a^*_1,\ldots,\delta a^*_L)^T$
the deviation from an arbitrary trajectory ${\bf a}(t)$ and linearizing the
DNLSE around this trajectory we have
\begin{equation}
\label{41}
i\,\frac{d}{dt}\delta{\bf a}= M\left[{\bf a}(t)\right]\, \delta {\bf a} \;,
\end{equation}
where $M\left[{\bf a}(t)\right]$ is a $2L\times 2L$ matrix of the following structure:
\begin{eqnarray}
\label{42}
M\left[{\bf a}(t)\right]=\left(
\begin{array}{cc}
A+gB&gC\\-gC^*&-(A+gB)^*
\end{array}\right) \; \\
A_{l,m}=-\frac{J}{2}\left(\delta_{l+1,m}e^{iFt}
+\delta_{l-1,m}e^{-iFt} \right) \;, \\
B_{l,m}=|a_l(t)|^2\delta_{l,m} \;,\quad
C_{l,m}=a^2_l(t)\delta_{l,m} \;.
\end{eqnarray}
Inserting the trajectory (\ref{40}) into (\ref{41}), the linear equation (\ref{41}) 
takes a form where the matrix $\widetilde{M}(t)$ is periodic in time. Finally, we 
introduce the stroboscopic map for the discrete time $t_n=T_Bn$: 
\begin{equation}
\label{43}
\delta\tilde{\bf a}(t_{n+1})=U\delta\tilde{\bf a}(t_{n}) \;,\quad
U=\widehat{\exp}\left(-i\int_0^{T_B}\widetilde{M}(t) dt\right) \;.
\end{equation}
This stability matrix $U$ is symplectic and hence the considered 
periodic trajectory (\ref{40}) is stable if and only if all its eigenvalues 
lie on the unit circle. 

Using the unitary transformation $U\rightarrow VUV^{-1}$, where
\begin{equation}
\label{44}
V=\left(
\begin{array}{cc}
T&0\\0&T
\end{array}\right) \; ,\quad
T_{l,m}=L^{-1/2}\exp\left(\frac{i2\pi k}{L}l\right) \;, 
\end{equation}
the stability matrix $U$ can be factorized into $L$ decoupled $2\times2$ 
matrices which can be labeled as $U^{(k)}$ with $k=0,\pm 1,\ldots,\pm(L-1)/2$. 
The matrix $U^{(0)}$ is not of 
interest here because its eigenvalues are always located on the unit circle. 
The explicit 
form of the matrix $U^{(\pm k)}$ is given by Eq.~(\ref{6}) and, depending 
on $F$, its eigenvalues $\lambda_{1,2}$ either lie on the unit circle 
(with $\lambda_2=\lambda_1^*$) or on the real axis (with $\lambda_2=1/\lambda_1$).
We have stability in the first, and instability in the second case.\\

\section{Mean-Field Ensemble}
\label{sec-ensemble}
Let us recall that the mean-field dynamics of the site
amplitudes $a_l$ (or, alternatively, the quasimomentum amplitudes
$b_k$) appears as a canonical
Hamiltonian evolution in a $2L$-dimensional complex phase space which
conserves the norm, i.e. the phase space is the surface of a complex
sphere. In order to approximate the many-particle dynamics, we construct an   
ensemble of mean-field trajectories representing the initial
many-body state $|\Psi(0)\rangle$. This is achieved most conveniently
by using the quantum Husimi phase space distribution \cite{07phase}, the 
projection onto $SU(L)$ coherent states which in the quasimomentum representation 
is given by
\begin{equation}
\label{cohstate} 
|{\bf b}\rangle_q=\frac{1}{\sqrt{N!}}\,\Big(\sum_{k=-(L-1)/2}^{(L-1)/2}b_k
\hat b_k^\dagger\Big)^N\,|0\rangle
\end{equation}
with ${\bf b}=(b_{-(L-1)/2},..,b_0,..,b_{(L-1)/2})$ and $\sum_k|b_k|^2=L$.
For the initial many-particle state (\ref{31})  this
Husimi density $Q({\bf b})=|\langle{\bf b}|\Psi(0)\rangle|^2$  is easily 
calculated as
\begin{equation}
Q({\bf b})= B |b_0|^{2N},
\label{ensemble2}
\end{equation}
where $B$ is a normalization constant.
Note that the probability density for the vector components depends only
on the zero-mode probability $|b_0|^2$ and is strongly localized in the
region close to its maximum value $|b_0|^2=L$ for large particle number $N$.

Numerically, one can construct the ensemble (\ref{ensemble2}) by means of a
rejection method \cite{Pres07}: First one generates $L$ randomly distributed 
real numbers $r_k$ in the unit 
interval with random phases
$\phi_k$ and normalizes the random vector with components $z_k=r_k\,e^{i\phi_k}$ 
to unity. Another random number $v$ in the unit interval is chosen in order to 
decide if the generated vector is accepted as a part of the ensemble: 
it is accepted if $v<|z_0|^{2N}$ and rejected otherwise. 
Finally we renormalize as $b_k=\sqrt{L}z_k$.
If desired, the ensemble can be Fourier transformed to yield the 
corresponding ensemble of lattice site amplitudes $a_l$.
%


\begin{thebibliography}{10}
\bibitem{remarkmf1}
In the numerical computations discussed in the following we used in most cases
  $N=100$ particles. The results are stable when $N$ is increased keeping the
  value of $WN$ constant unless explicitely stated.
\end{thebibliography}

\end{document}